\documentclass[prb,aps,twocolumn,amsmath,amssymb,superscriptaddress]{revtex4-2}
\usepackage{graphicx,picture,calc}
\usepackage{verbatim}
\usepackage{braket}
\usepackage{epsfig}
\usepackage{epstopdf}
\usepackage{amsfonts}
\usepackage{amsthm}
\usepackage{amsmath}
\usepackage{amssymb}
\usepackage{color}
\usepackage[usenames,dvipsnames,svgnames,table]{xcolor}
\usepackage{hyperref} 
\hypersetup{colorlinks=true,linkcolor=NavyBlue,citecolor=BrickRed,urlcolor=NavyBlue}
\usepackage{dsfont}
\usepackage{color}
\usepackage{grffile}
\usepackage{bm}

\newcommand{\be}{\begin{equation}}
\newcommand{\ee}{\end{equation}}

\newcommand{\bea}{\begin{eqnarray}}
\newcommand{\eea}{\end{eqnarray}}
\newcommand{\beal}{\begin{align}}
\newcommand{\eeal}{\end{align}}

\begin{document}

\title{Nematic order driven by superconducting correlations}
\author{Finn Lasse Buessen}
\email{fbuessen@physics.utoronto.ca}
\affiliation{Department of Physics, University of Toronto, Toronto, Ontario M5S1A7, Canada.}
\author{Sopheak Sorn}
\email{ssorn@physics.utoronto.ca}
\affiliation{Department of Physics, University of Toronto, Toronto, Ontario M5S1A7, Canada.}
\author{Ivar Martin}
\email{ivar@anl.gov}
\affiliation{Materials Science Division, Argonne National Laboratory, Lemont, IL  60439, USA}
\author{Arun Paramekanti}
\email{arunp@physics.utoronto.ca}
\affiliation{Department of Physics, University of Toronto, Toronto, Ontario M5S1A7, Canada.}

\date{\today}

\begin{abstract}
The interplay of nematicity and superconductivity has been observed in a wide variety of quantum materials. To explore this interplay,
we consider a two-dimensional (2D) array of nematogens, local droplets with $Z_3$ nematicity, coupled to a network of Josephson junction wires.
Using finite temperature classical Monte Carlo simulations, we elucidate the phase diagram of this model and show that the development of
superconducting correlations and the directional delocalization of Cooper pairs can promote nematogen ordering, resulting in long-range nematic order. 
We obtain the transport properties of our model within an effective 
resistor network picture.  We discuss these ideas in the context of the 2D electron gas at the (111) KTaO$_3$ interface and the doped
topological insulators Nb$_x$Bi$_2$Se$_3$ and Cu$_x$Bi$_2$Se$_3$. Our work makes contact with
Phil Anderson's numerous contributions to broken symmetries driven by the saving of kinetic energy, 
including double exchange ferromagnetism and the interlayer tunneling theory of high $T_c$
superconductivity.
\end{abstract}

\maketitle


\section{Introduction}

The electron nematic \cite{NematicReview2010}, a liquid crystalline state of electrons which exhibits spontaneous breaking of lattice rotational symmetry, has been extensively explored in quantum Hall systems~\cite{Koulakov_PRL1996,Du_SSC1999,Lilly_PRL1999,Fradkin_PRB1999,Abanin_PRB2010,JXia_NPhys2011,Feldman_Science2016,Sodemann_PRX2017}, correlated superconductors~\cite{Tranquada1995,ChenFang2008,kruger2009,Lawler2010,Fernandes2012,Billinge2014,Gastiasoro2014,Nie2014,Hosoi2016,Nie2017,Paglione2020}, and the bilayer ruthenate compound Sr$_3$Ru$_2$O$_7$~\cite{Kee2005,Borzi2007,Raghu2009,Green2012,Hayden2015}.
In these systems, nematic order emerges as a vestige \cite{fernandesARCMP,Nie2014,Nie2017} of underlying spin or charge density wave orders, or due to a density imbalance between orbital or valley degrees of freedom.
Quantum fluctuations in the nematic order can potentially act as a pairing glue for electrons, resulting in a purely electronic mechanism for superconductivity (SC)~\cite{Lederer2015,Klein2018}. 
This idea has been substantiated using sign-problem-free quantum Monte Carlo (MC) simulations of electrons coupled to a quantum Ising model of nematic order ~\cite{Lederer2017,bergARCMP}.

\begin{figure}[t]
\includegraphics[width=\linewidth]{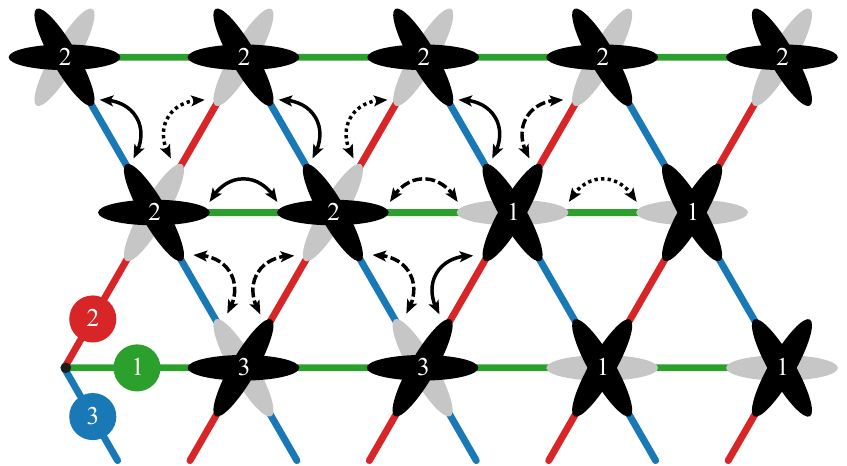}
\caption{Lattice model of $Z_3$ nematogens. 
Nematogens are depicted with black/gray lobes, with orientations labeled by their $Z_3$ states $\theta_i=1,2,3$. 
They are coupled by three sets of Josephson junction wires (green:1, red:2, blue:3), oriented along the three lattice directions. 
The intra-wire nearest-neighbor Josephson coupling $\propto \! J_h$ is modulated by the orientation of the two adjacent nematogens; the three different potential nematogen combinations coupled on a bond (black/black, black/gray, gray/gray) are indicated by (solid, dashed, dotted) arrows. 
In addition, the three emanating wires at each site are coupled by an onsite Josephson coupling $J_\ell$.}
\label{fig:model}
\vskip -0.3cm
\end{figure}

In this paper, we consider a model of nematogens, droplets with local $Z_3$ nematicity, coupled to 
one-dimensional (1D) Josephson 
junction wires (JJWs), as shown in Fig.~\ref{fig:model}. The Cooper pair hopping  between adjacent sites along a given direction is controlled by their local nematogen orientations. We show that this model realizes a converse
scenario where superconducting correlations -- instead of being driven by nematicity --  are responsible for establishing nematic 
order in the first place.

Our work is partly motivated by the discovery of SC and nematicity in the doped topological insulators Nb$_x$Bi$_2$Se$_3$ and Cu$_x$Bi$_2$Se$_3$~\cite{FuBerg2010,Matano2016,Maeno2017,Lortz2020}. Doped Bi$_2$Se$_3$ has been proposed to host a 
two-component superconducting order parameter $(\psi_1,\psi_2)$,  transforming as a two-dimensional 
irreducible representation of the crystalline point group symmetry \cite{FuBerg2010,Matano2016,Maeno2017,Lortz2020}. This 
admits time-reversal breaking SC with $(\psi_1,\psi_2) \equiv \psi_0 (1, \pm i)$ as well as nematic pairing states with
$(\psi_1,\psi_2) \equiv \psi_0 (\cos\theta,\sin\theta)$, where the crystalline $C_3$ symmetry dictates $\theta=0,2\pi/3,4\pi/3$.
A recent experiment favors the latter possibility, with the observation of 
$Z_3$ nematic order below $T_n \!\approx\! 3.8$\,K, while SC only occurs at a lower temperature scale $T_c\!\approx\!3.25$\,K~\cite{Lortz2020}.
In addition, there is evidence for diamagnetism already near $T_n$, hinting at SC fluctuations being important near and above the nematic transition~\cite{Lortz2020},
and experiments have demonstrated uniaxial strain control of nematic domains \cite{kostylev2020}. It is thus plausible that there are nematogens
even above $T_n$ in doped Bi$_2$Se$_3$, which correspond to
domains of a phase-fluctuating nematic pairing state with different nematic orientations.

A similar interplay of nematicity and SC has also been observed in a 2D electron gas (2DEG) formed at (111) KTaO$_3$ oxide interfaces~\cite{KTO_liu2020}. 
At a carrier density $n\! \sim\! 10^{14}$/cm$^2$, SC occurs in the 2DEG with
$T_c \! \approx \! 2$\,K \cite{KTO_liu2020,KTO_chen2020,Ma_2020}, but an underlying nematicity is revealed via an anisotropy in the 
non-linear current-voltage characteristics beyond the critical current density \cite{KTO_liu2020}.
At a lower density $n \!\sim\! 3 \! \times \!10^{13}$/cm$^2$, the nematicity manifests itself already in the normal state, as an
anisotropic resistivity below 
$T_n \!\approx \! 2$\,K, before SC sets in at $T_c\approx 0.5$\,K~\cite{KTO_liu2020}.
Recent work has proposed that this anisotropy might stem from spin-stripe order due to a nested hexagon-shaped Fermi surface (FS) \cite{arribi2020}, 
with SC being stabilized on those faces of the hexagonal FS which remain ungapped by stripe order.
However, transport below $T_n$ in the low density 2DEG exhibits a remarkable non-mean-field temperature dependence: the resistivity increases along one 
direction, $[1\bar{1}0]$, as expected for gapping of a Fermi surface, but it \emph{decreases} by a similar amount along the orthogonal $[11\bar{2}]$ direction~\cite{KTO_liu2020}.
This behavior is instead reminiscent of a resistor network made of preformed nematogens which act as anisotropic resistance units.
In the KTaO$_3$ 2DEG, with underlying $C_3$ symmetry,
the nematogens may thus correspond to $Z_3$ domains of unidirectional spin stripe order coexisting with SC \cite{arribi2020}.
In this scenario, if the resistor network is disordered, it leads to an average isotropic resistivity. 
Long-range order of the nematogens, on the other hand, naturally results in an increased resistance along one direction and a decreased 
resistance in the transverse direction. In KTaO$_3$,
the experimental observation \cite{KTO_liu2020} that normal state nematicity  does not extend far above the superconducting 
$T_c$ in zero magnetic field, or exists only in a small window above the critical field for $T < T_c$, 
again suggests that local superconducting correlations are likely to be important in establishing nematic order. 
Consistent with this scenario, an in-plane field which is less effective at suppressing SC also has a smaller impact on the nematic resistivity \cite{KTO_liu2020}.

We thus propose that the experiments on doped Bi$_2$Se$_3$ and the KTaO$_3$ 2DEG may be fruitfully viewed in terms of 
preformed mesoscopic $Z_3$ nematogens, with their ordering being driven by the Josephson coupling between nematogens and the
resulting delocalization of Cooper pairs. Such symmetry
breaking driven by a ``kinetic energy saving'' mechanism is reminiscent of the Zener and Anderson-Hasegawa theory for
the establishment of ferromagnetism in double exchange magnets, where local moments order due to the lowering of kinetic energy by
electron delocalization \cite{DE1,DE2}. A similar mechanism of saving Cooper pair
kinetic energy provided the impetus for the interlayer tunneling theory  (ILT) by Anderson and coworkers
\cite{ILT1,ILT2,ILT3}, yielding a mechanism for the emergence of high temperature 3D SC from coupling of 2D non-Fermi liquids. The ILT
was later found to be in conflict with experiments on the cuprates, but a spin analogue of ILT
provides an excellent description of the quasi-1D frustrated magnetism in Cs$_2$CuCl$_4$ \cite{Kohno2007}.

Our theory may also be viewed as a bosonic variant of 
the Kugel-Khomskii model \cite{Kugel1982,Aharony2005} which describes spin and  orbital order of electrons in solids.
In this analogy, SC and $Z_3$ nematicity
act respectively as `spin' and `orbital' degrees of freedom at a site, while inter-site and local Josephson couplings on the JJWs play the role of
orbital-dependent exchange interaction and Hund's coupling respectively.

 \section{Landau theory}
 To study the interplay of superconductivity and nematicity, we begin with a conventional Landau theory description. We 
 introduce two complex order parameters: 
 $\Phi$ for SC, and $\Psi$ for the $Z_3$ nematic. The independent order parameters have Landau free energies
\bea
S_{\Phi} &=& \int d^2 r \left[ r_s |\Phi|^2 + u_s |\Phi|^4 + \kappa_s |\vec\nabla \Phi|^2 + \ldots \right]\label{s1}\\
S_{\Psi} &=& \int d^2 r  \big[  r_n |\Psi|^2 + w_n (\Psi_n^3 + \Psi_n^{*3}) + u_n |\Psi|^4 \notag\\
&+& \kappa_n |\vec\nabla \Psi|^2+ \ldots \big],  \label{s2}
\eea
where the cubic term is permitted by $C_3$ rotation symmetry under which $\Psi \to e^{i4\pi/3}\Psi$. This reduces the nematic theory to a $3$-state clock/Potts model.
The coupling between the nematic order and the SC takes the form
\bea
\!\!\!\!S_{\Phi,\Psi} &\!=\!&\! \int \!\! d^2 r \! \left[u_{sn}  |\Phi|^2 |\Psi|^2 \!+\!  \kappa_{sn} \{\Psi ~ \Phi^* \partial^2_{+} \Phi \!+\! {\rm c.c.} \} 
\!+\! \ldots \right]
\label{s3}
\eea
where $\partial_{\pm} \equiv \partial_x \pm i \partial_y$ and ``c.c'' refers to complex conjugate. The gradient coupling is chosen to be invariant 
under a $C_3$ rotation which leads to 
$\Psi \to \Psi e^{i 4\pi/3}$ and $\partial_{\pm} \to \partial_{\pm} e^{\mp i 2 \pi/3}$. We have omitted additional gradient terms, and higher order terms in this action. 
The gradient coupling $\kappa_{sn}$ 
causes the superconducting stiffness to become anisotropic in the presence of nematic order $\langle \Psi \rangle \neq 0$. At the same time, integrating out the fluctuating superconducting
order parameter $\Phi$ from the gradient terms can renormalize the nematic mass $r_n$ and stiffness  $\kappa_n$, thus helping to stabilize nematic order. 

As an illustration, we compute the mass renormalization to Gaussian order by dropping 
$u_s$ and keeping $r_s > 0$ in Eq.~\ref{s1}. Ignoring $u_{sn}$ in Eq.~\ref{s3}, and integrating out $\Phi$ leads to
\bea
\tilde{r}_n = r_n - \kappa_{sn}^2 \int_0^\Lambda \! d^2 q~ \frac{q^4}{(r_s+\kappa_s q^2)^2}.
\eea
As $r_s$ decreases, the superconducting correlation length grows, and the renormalized nematic mass can change sign, 
$\tilde{r}_n < 0$, thus favoring nematic order, even when the bare $r_n > 0$.

In the (111) KTaO$_3$ 2DEG, we may view $\Psi$ as a vestige of spin-density stripe order, with the SC order described by $\Phi$. In this, we are
assuming that the SC order parameter in KTaO$_3$ is a conventional single-component order parameter. In doped 
Bi$_2$Se$_3$,  the two-component nematic superconducting order parameter may be expressed as a composite: 
$(\psi_1,\psi_2) \equiv \Phi ({\mathrm Re} \Psi, {\mathrm Im} \Psi)$. We note that recent work has proposed a distinct Landau theory for 
nematic SC in Bi$_2$Se$_3$ \cite{fernandes2021charge4e} which highlights its connection to charge-$4e$ SC 
\cite{fernandes2021charge4e,jian2021charge4e,zeng2021phasefluctuation}.


\section{Lattice model}

In order to study the interplay of SC and nematic order beyond Landau theory, and in view of the aforementioned experiments, we consider a model of $Z_3$~nematogens coupled to a network of JJWs, as schematically depicted in Fig.~\ref{fig:model}. 
The Hamiltonian is given by
\begin{equation}
\!\!H \!=\! - J_h \!\! \sum_{i,\mu} \!\cos(\varphi^\mu_i-\varphi^\mu_{i+\mu}) \Delta_{i,i+\mu}  \! -\! J_{\ell}\!\!\!\! \sum_{i,\mu < \nu}\!\!\!  \cos(\varphi^\mu_i-\varphi^{\nu}_i) \,,
\label{eq:H}
\end{equation}
where $\varphi^\mu_i$ denotes the superconducting phase at site $i$ for wire $\mu \!=\! 1,2,3$. 
The first term $J_h>0$ denotes Cooper pair hopping between sites $i$ and $i\!+\!\mu$, with the latter being the nearest neighbor of site $i$ along wire $\mu$, and the second term $J_\ell$ is the local Josephson coupling between the three wires meeting at each site. 
The information about the nematogen configuration is contained in the directional Josephson coupling defined as
\begin{equation}
\Delta_{i,i+ \mu}^{-1} \! = \! g_{i,i+\mu}^{-1} + g_{i+\mu,i}^{-1} \,,
\label{eq:josephsoncoupling}
\end{equation}
where the conductance $g_{i,i+\mu}$ depends on the $Z_3$ nematogen orientation $\theta_i \!=\! 1,2,3$ as
\begin{equation}
g_{i,i+\mu}= \begin{cases}
    1-\eta & \text{if } \theta_i=\mu\\
    1+\eta/2     &     \text{if } \theta_i \neq \mu\\
    \end{cases} \,.
\label{eq:g}
\end{equation}
Following these definitions, for $0 \!<\!  \eta \! < \! 1$, a nematogen with configuration $\theta_i\!=\! \mu$ suppresses the Josephson coupling along wire $\mu$ relative to the other two directions; on the other hand, for $-2 \!<\! \eta\!<\! 0$, 
the $\theta_i\!=\! \mu$ configuration enhances the Josephson coupling along wire $\mu$.
Depending on the orientations of the two adjacent nematogens, the nearest-neighbor Josephson coupling $\Delta_{i,i+\mu}$ can take on three values: $\Delta_{bb}$, $\Delta_{bg}$, or $\Delta_{gg}$, where the subscripts denote the colors of the lobes (b=black, g=gray) pointing towards each other as illustrated in Fig.~\ref{fig:model}.
They assume explicit values $\Delta_{bb}=(2+\eta)/4$, $\Delta_{gg}=(1-\eta)/2$, and $\Delta_{bg}=(\eta+2)(1-\eta)/(4-\eta)$, respectively.

\begin{figure}
\includegraphics[width=\linewidth]{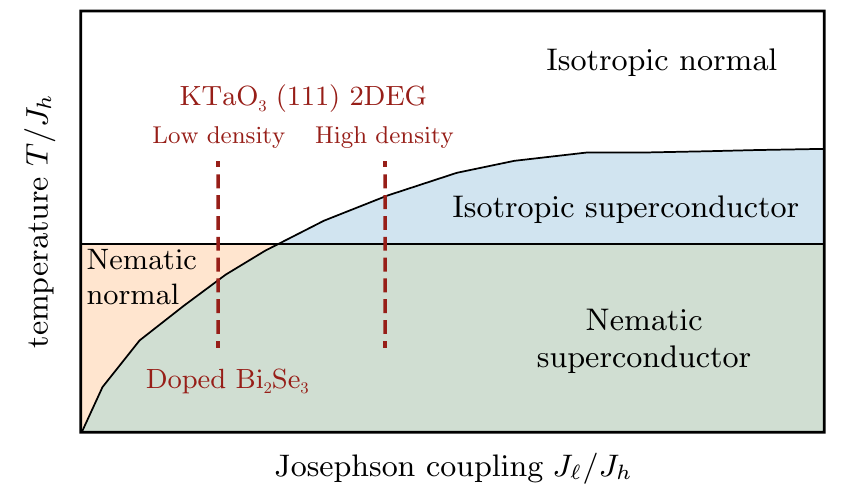}
\caption{Schematic phase diagram for the model Hamiltonian as a function of the on-site Josephson coupling $J_\ell/J_h$ and temperature $T/J_h$ at fixed $\eta$. 
The high-temperature isotropic normal state is connected to the low-temperature nematic superconductor either via an intermediate nematic normal phase or an isotropic superconductor. 
Dashed lines indicate cuts relevant to low-density and high-density KTaO$_3$ (111) 2DEG, and to doped Bi$_2$Se$_3$.}
\label{fig:schematic}
\vskip -0.3cm
\end{figure}

In the limit $J_\ell\!=\!0$, the different JJWs are decoupled from each other, and we can focus on an individual JJW with periodic boundary conditions which couple chains of nematogens.
For $\eta\! >\! 0$, we find $\Delta_{bb} \!>\! \Delta_{bg} \!>\! \Delta_{gg}$. 
Minimizing the Josephson coupling energy on a single JJW (say, green:1) only requires that the nematogen at each site is constrained to be $\theta_i \!\neq\! 1$, yielding $\Delta_{i,i+1}\!=\! \Delta_{bb}$ for every pair of nearest neighbors.
However, if we consider adjacent parallel wires, it is easy to check from Fig.~\ref{fig:model} that the lowest energy for a system with periodic boundary conditions is achieved only when all the nematogens in both wires are globally aligned. 
Consequently, the ground states of the full 2D model will also exhibit nematic order. 
In a similar fashion, one can establish ground state nematic order for $\eta \!<\! 0$.
We emphasize that, since the nematic order is a discrete order, it can remain stable at finite temperature in the thermodynamic limit. 
However, since the individual JJWs remain decoupled 1D XY-type wires, there is no SC for any $T > 0$; we thus expect a nematic ordering transition at $T_n\! \propto \! J_h$ and a superconducting transition temperature $T_c\!=\! 0$. 
Next, when we switch on weak onsite Josephson coupling $0 \!<\! J_\ell \! \ll \! J_h$, global 2D SC is established with $T_c \!<\! T_n$, leading to a window of normal state nematic order at intermediate temperatures.
On the other hand, as $J_\ell \!\to\! \infty$, the phases $\varphi_i^\mu$ on different wires get locked at each site, leading to a single triangular lattice JJ array, for which $T_c$ may be larger than $T_n$.
These considerations lead to the schematic phase diagram for the model Hamiltonian Eq.~\eqref{eq:H} as a function of $J_\ell/J_h$ and temperature (for fixed $\eta$) shown in Fig.~\ref{fig:schematic}. 
Below, we confirm the phase diagram using classical MC simulations. We note that the nematic transition temperature $T_n$ is nearly independent of the {\it onsite}
Josephson coupling, but is instead dictated by the intersite Josephson coupling $J_h$ since it is driven by Cooper pair hopping.

\begin{figure*}
\includegraphics[width=\linewidth]{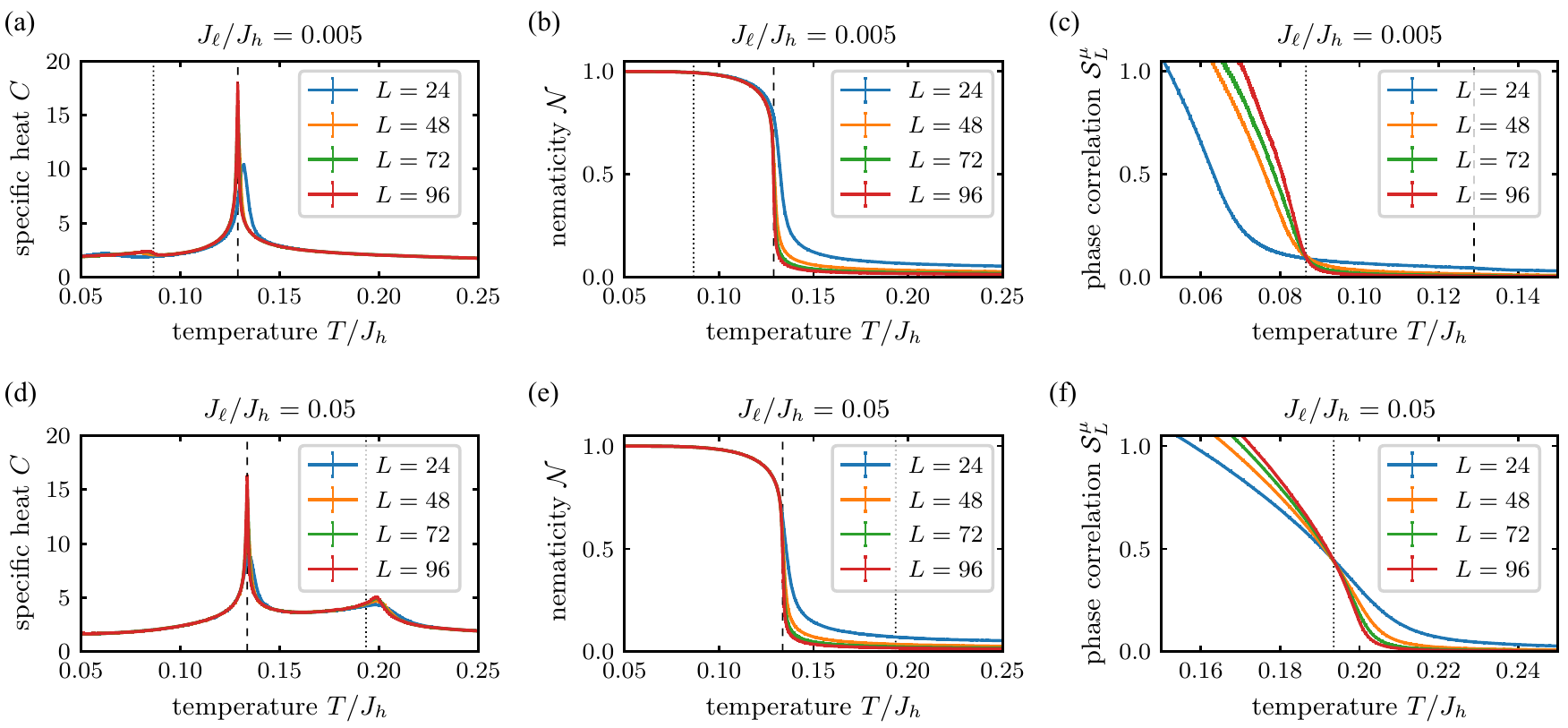}
\caption{Monte Carlo simulation results for the model Hamiltonian in Eq.~\eqref{eq:H}, for fixed $\eta\!=\!2/3$ at (a-c) $J_{\ell}/J_h\!=\!0.005$ and (d-f) $J_{\ell}/J_h\!=\!0.05$, respectively. 
(a)~Specific heat per site showing a sharp peak at the nematic transition at $T_n \! \approx \! 0.13 J_h$ (dashed line), and a weak bump near the BKT superconducting transition $T_c \!\approx \! 0.09 J_h$ (dotted line). 
(b)~The nematic order parameter~$\mathcal{N}$ near the nematic transition point. 
(c)~The scaled superconducting phase correlator~$\mathcal{S}_L^\mu$ shows a system-size independent crossing point at the superconducting BKT transition. 
Panels~(d-f): Similar results, but for $J_\ell/J_h\!=\!0.05$, showing a nearly unchanged $T_n$ but a significantly higher $T_c \!\approx\! 0.19 J_h \! > T_n$.}
\label{fig:mcplots}
\end{figure*}


\section{Monte Carlo study}

We have carried out classical finite-temperature MC simulations of the model defined in Eq.~\eqref{eq:H}. We studied finite systems of $L\!\times\!L$ unit cells with periodic boundary conditions and system sizes up to $L\!=\!96$. We equilibrated the system for $5\times 10^5$ MC sweeps, where a single sweep is defined as one attempted update per degree of freedom on average, before taking measurements for up to $5\times 10^6$ sweeps. 
For an improved sampling of the configuration space -- in particular when resolving the Berezinskii Kosterlitz Thouless (BKT) transition \cite{Berezinskii1971,KT1973} in a nematic background -- we implemented a parallel tempering scheme across 196 temperature points in an optimized temperature ensemble in the range $0.05<T/J_h<0.25$~\cite{Hukushima1996,Katzgraber2006}. 

We explore the phase diagram as a function of $J_{\ell}/J_h$ and temperature, keeping $\eta\!=\! 2/3$ fixed. 
In order to discriminate phases and detect phase transitions, we compute the specific heat, a nematic order parameter, and superconducting phase correlations on the JJWs.
The nematic order parameter is defined as $\mathcal{N} \!=\! (1/L^2) \langle \big| \sum_i e^{i 2 \pi \theta_i/3} \big| \rangle$, where $\theta_i\!=\! 1,2,3$ labels the
local nematogen configuration; 
the superconducting phase correlations are defined as $\mathcal{S}^\mu \!=\! (1/L^4) \langle \big| \sum_i e^{i \varphi^\mu_i} \big|^2 \rangle$. 
The latter quantity is particularly useful for the following reason. When $J_\ell \!>\! 0$, we expect the superconducting transition to be a BKT transition, which implies a universal $r^{-1/4}$ power law decay of phase correlations at the critical point.
This power law manifests in the finite-size dependence of the phase correlations, $\mathcal{S}^\mu \sim L^{-1/4}$. 
As a result, the scaled superconducting phase correlations $\mathcal{S}^\mu_L = L^{1/4}\mathcal{S}^\mu$ are expected to be system-size independent at the BKT critical point, and
the $\mathcal{S}^\mu_L$ curves for different $L$ should cross
at the BKT transition temperature \cite{XLi2014}.

The results of our MC analysis are summarized in Fig.~\ref{fig:mcplots}. 
Panels \ref{fig:mcplots}a and \ref{fig:mcplots}d show the specific heat for $J_\ell/J_h\!=\! 0.005$ and $J_\ell/J_h\!=\! 0.05$, respectively. In both cases, we observe a sharp peak at $T_n \! \approx\! 0.13 J_h$ which we associate with the onset of nematicity. 
This is supported by the nematic order parameter becoming finite at the same temperature scale (Figs.~\ref{fig:mcplots}b and \ref{fig:mcplots}e). 
The BKT transition is detected as a crossing point of the curves of $\mathcal{S}^\mu_L$ for different system sizes, as shown in Figs.~\ref{fig:mcplots}c and \ref{fig:mcplots}f. 
We point out that at a BKT transition, it is well known that there is an undetectable essential singularity in the specific heat; however, a rough indication of its location is given by a weak bump in the specific heat associated with the quenching of entropy tied to phase fluctuations. 

We distinguish two qualitatively different cases. For $J_\ell/J_h\!=\!0.005$, the superconducting $T_c \!\approx\! 0.09 J_h$ lies below the nematic transition temperature $T_n$. 
For $J_\ell/J_h\!=\!0.05$, on the other hand, $T_{c} \!\approx\! 0.19 J_h$, so that $T_c \!> \! T_n$. 
The existence of these two different sequences of phase transitions, with an intermediate phase which is either nematic with finite resistance or isotropic and superconducting, confirms the schematic phase diagram in Fig.~\ref{fig:schematic}. 


\section{Transport} 
\begin{figure}[t]
	\centering
	\includegraphics[width=0.4\textwidth]{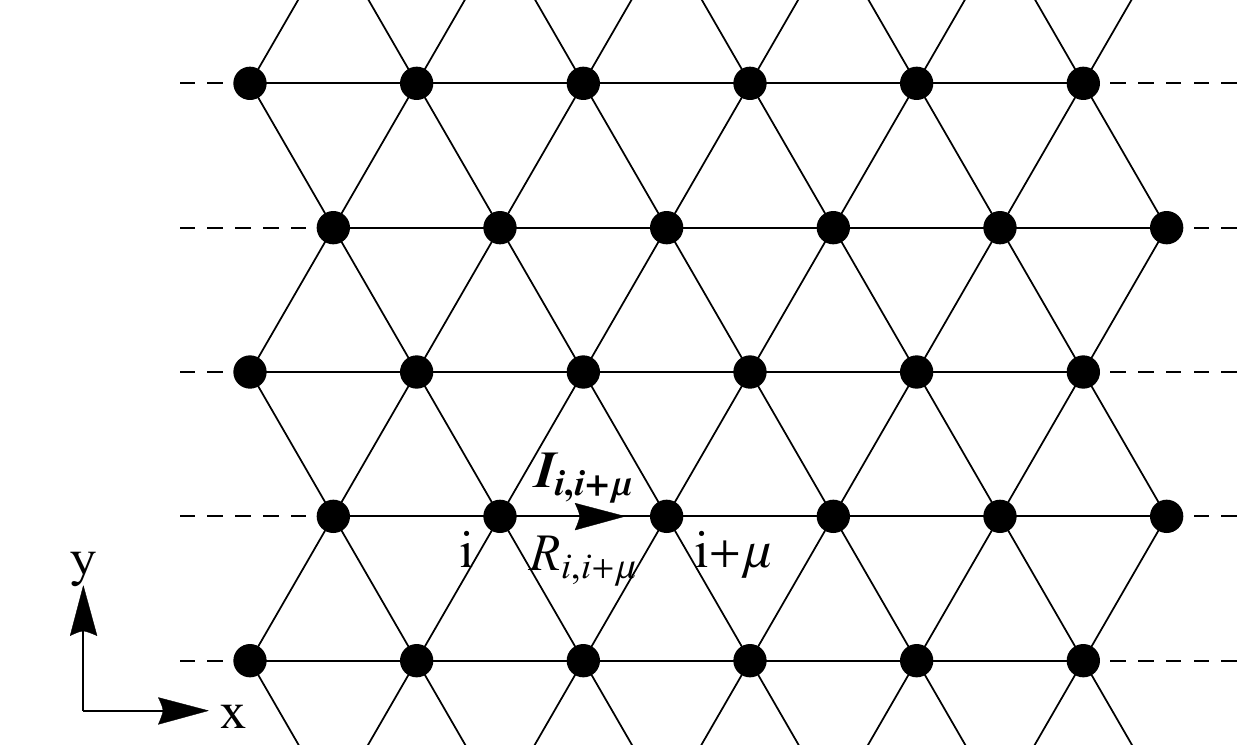}
	\caption{Triangular lattice of resistors residing on each bond whose resistance $R_{i,i+\mu}$ is proportional to $\Delta^{-1}_{i, i+\mu}$. Sites on the left edge are held at the same potential, and so are those on the right. A potential difference between the edges induces current $\mathbf{I}_{i,i+\mu}$ which is a vector quantity. The y-direction is periodic, and the dashed lines represent wires with zero resistance.}
	\label{fig:resistor_network}
\end{figure}

In the normal state, far above $T_c$, superconducting correlations are short ranged, and the inter-site Josephson links act as normal resistances $\propto \!  \Delta_{i,i\pm \mu}^{-1}$.
We can thus approximately compute transport properties in the normal state by translating nematogen configurations into configurations of a corresponding resistor network~\cite{RN1,RN2}. 
Such a resistor network is shown in Fig.~\ref{fig:resistor_network}, where
we consider a triangular mesh of sites $\{i\}$ with a resistor on each nearest-neighbour bond \cite{RN1} whose resistance is proportional to $\Delta_{i, i+\mu}^{-1}$. 
To obtain the effective resistivity ~\cite{RN1, RN2} for a given configuration $\{\Delta_{i, i+\mu}\}$, we apply a potential difference between the two edges, as illustrated in Fig.~\ref{fig:resistor_network}, with periodic boundary conditions along y, and solve equations from Kirchhoff's laws to obtain the currents $\mathbf{I}_{i,i+\mu}$. The current densities $j_x$ and $j_y$ are computed and used to determine the conductivities $\sigma_{xx}$ and $\sigma_{yx}$. Similarly,
$\sigma_{yy}$ and $\sigma_{xy}$ can be obtained similarly by exchanging the roles of the periodic and open boundaries. The conductivity tensor is then inverted and diagonalized to arrive 
at the principal eigenvalues of the resistivity tensor.
We finally average these results over $100$ nematogen configurations drawn from our MC simulations at each temperature. 

\begin{figure}[b]
\includegraphics[width=\linewidth]{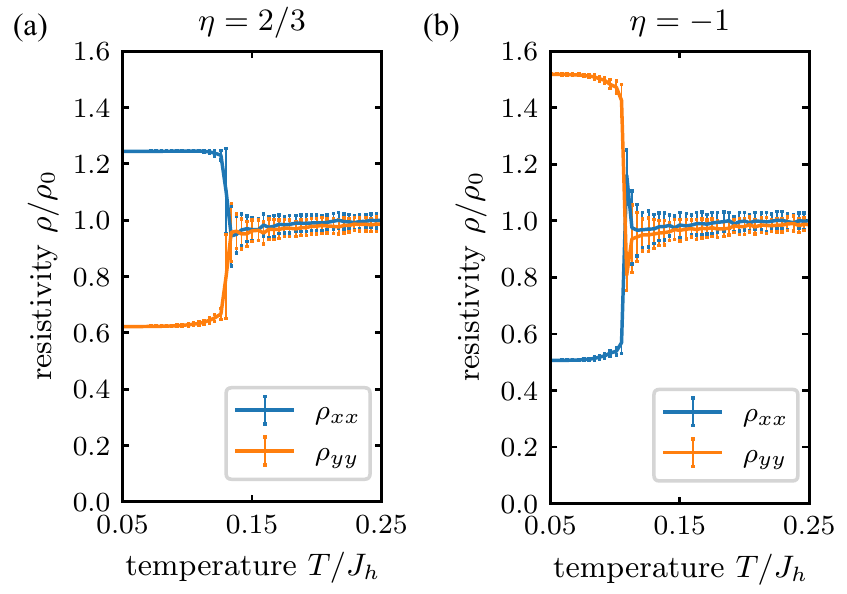}
\caption{Resistivity tensor eigenvalues for  the classical resistance network model with (a) $\eta=2/3$ (hard axis along the x-direction) and (b) $\eta=-1$ 
(easy axis along the x-direction).
The network is obtained by using conductances as given in Eq.~\eqref{eq:josephsoncoupling}, with nematogen configurations drawn from the MC simulations of the model in Eq.~\eqref{eq:H} with $J_{\ell}\!=\!0.005$, on an $L\!\times\! L$ system with $L\!=\!20$. We average the resistivity 
over $100$ configurations and normalize it by the high temperature isotropic value $\rho_0$.}
\label{fig:res}
\vskip -0.3cm
\end{figure}

Fig.~\ref{fig:res} shows the eigenvalues of the resistivity tensor, which correspond to $\rho_{xx}$ and $\rho_{yy}$ if we choose the unique hard (easy) axis in the nematic phase to be along the $x$-direction, corresponding to the parameter choice $\eta > 0$ ($\eta < 0$).
As we cool below $T_n$, the resistivity increases along one direction and decreases along the other direction, consistent with a symmetry analysis discussed below.
This anisotropic behavior of the resistivity in the nematic normal state is in qualitative agreement with the experiments on the KTaO$_3$ (111) 2DEG.

We interpret our findings in terms of the nematic order parameter $\Psi = |\Psi| e^{i\theta}$, in terms of which 
the change in the resistivity tensor due to nematicity
takes on the following symmetry-dictated form:
\be
\Delta\rho \propto |\Psi| \begin{pmatrix} \cos\theta & \sin\theta\\ \sin\theta & -\cos\theta\end{pmatrix}
\ee
This form ensures that a $C_3$ rotation of the nematic order, which sends $\theta \to \theta + 4\pi/3$, can be encoded in a spatial rotation via
$\Delta\rho \to R^T \Delta\rho R$, where $R$ is the $2\times 2$ rotation matrix. The tensor $\Delta\rho$ has
eigenvalues $\pm |\Psi|$. This eigenvalue splitting reveals itself in Fig.~\ref{fig:res} as we go below the nematic transition. 
For $\theta=0$, this leads to $\Delta\rho_{xx} = - \Delta\rho_{yy}$ and $\Delta\rho_{xy}=\Delta\rho_{yx}=0$, while 
for $\theta=2\pi/3,4\pi/3$, there is a symmetric off-diagonal component to the resistivity tensor.

The classical resistor network model is a valid approach to compute the resistance of the Josephson junction array when superconducting correlations are very short-ranged. However, as we approach the BKT superconducting transition $T_c$, with $T_c \! \ll \!T_n$ when $J_\ell \!\ll\! J_h$, 
these superconducting correlations grow and must be taken into account. 
This will eventually lead to vanishing resistivity
along both directions at $T_c$. A phenomenological route to incorporating these correlations is to view patches of linear dimension $\xi(T)$, where $\xi(T)$ is the temperature dependent correlation length measured in units of the lattice constant, 
as zero-resistance `short' regions. We then expect the resulting network to have a 
renormalized resistivity
\bea
    \tilde{\rho}(T) &=& \rho(T) /\xi(T),
\eea
where the `bare resistivity' $\rho(T)$ is shown in Fig.~\ref{fig:res} for $J_\ell/J_h=0.005$.
\begin{figure}[t]
    \centering
    \includegraphics[width = 0.48\textwidth]{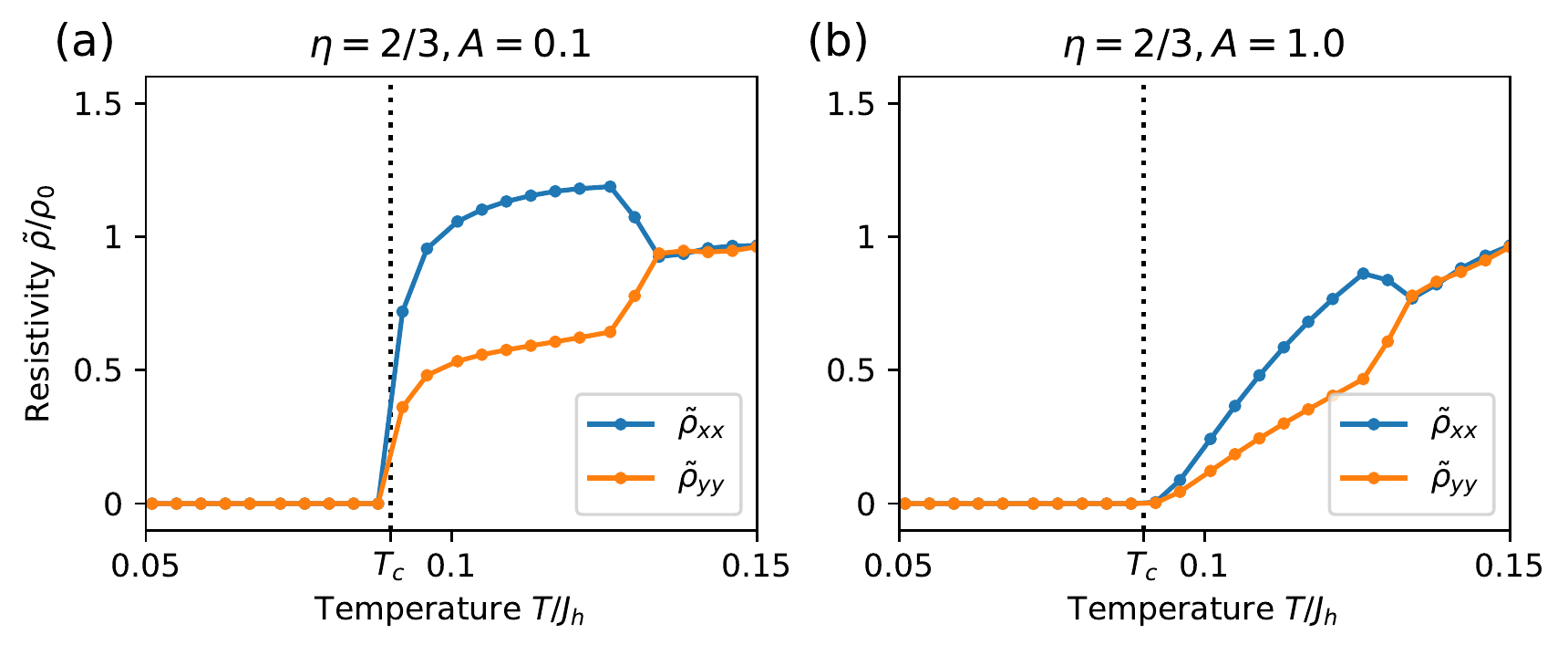}
    \caption{Illustration of the impact of growing superconducting correlations on the resistive anisotropy as a function of temperature.}
    \label{fig:renormalized}
\end{figure}

Fig.~\ref{fig:renormalized} shows the renormalized resistivity, which becomes zero below $T_c$, where we have used $\rho(T)$ 
obtained from the resistor network model, and the following ansatz for the BKT correlation length:
\bea
    \xi(T) &=&\exp\left[{\frac{A}{\sqrt{T/T_c - 1}} }\right].
\eea
Here,  $T_c = 0.09 J_h$ as obtained from our Monte 
Carlo simulations shown in Fig.~\ref{fig:res} for $J_\ell/J_h=0.005$. We choose two illustrative cases
$A = 0.1, 1.0$ to obtain the renormalized $\tilde{\rho}(T)$. These plots are in qualitative agreement with experimental data on the (111) KTaO$_3$ 2DEG.
In the opposite regime, when $T_c \!>\! T_n$ in the (111) KTaO$_3$ 2DEG, SC develops before the onset of nematicity. 
The superconductor should then exhibit an anisotropic critical current.
In doped Bi$_2$Se$_3$, it is possible that this window of anisotropic resistivity might be small.


\section{Impact of a magnetic field}

A perpendicular magnetic field will suppress the SC gap on individual grains, leading to a decrease in the Josephson couplings $J_h, J_\ell$. 
Within Landau theory, we expect $J_h(B)$ and $J_\ell(B)$ to decrease $\propto\! (1\!-\! B/B_c)$, where $B_c$ is the bulk upper critical field.
Since $T_n \propto J_h$, with $J_h \!\to\! 0$ marking the point where the driving force for nematicity is lost, we also expect $T_n(B) \!\propto\! (1-B/B_c)$. 
Here $B_c \!\sim\! \Phi_0/\xi_0^2$, with $\Phi_0$ and $\xi_0$ being the superconducting flux quantum and coherence length, respectively.
Experiments on the (111) KTaO$_3$ 2DEG at higher densities have found that bulk SC gets suppressed for $B_c \! \sim \! 1$\,T, implying
 $\xi_0 \!\sim\! 10$\,nm~\cite{KTO_liu2020}.

However, the perpendicular magnetic field may also lead to Josephson frustration, if we recognize that our  model of Josephson coupling between superconducting grains being mediated by 1D wires reflects a convenient idealization of the real system. 
In reality, the Josephson coupling between grains of size $\xi_g \gg \xi_0$ will occur via the entire inter-grain region. When the field is strong enough to insert a vortex in this region, it can effectively suppress the Josephson coupling between two adjacent grains. 
This interference effect, which leads to the familiar Fraunhofer-like pattern in Josephson junctions~\cite{Dynes1971,Robinson2019},
would manifest itself at a much smaller field scale $B_g \sim \Phi_0/\xi_g^2$ set by the grain size.
Assuming $\xi_g \!\sim\! 10 \xi_0$ would yield $B_g \!\sim\! 10$\,mT.
The published data on lower density (111) KTaO$_3$ 2DEG shows evidence of two distinct field scales and interference-like effects in magnetotransport~\cite{KTO_liu2020}, which may reflect both of these mechanisms being at work. 
A complete account of magnetotransport phenomena is a topic for future study.


\section{Impact of disorder and strain}

In our model, it is clear that short-range superconducting correlations are sufficient to establish 2D nematic order. 
Disorder which limits the range of superconducting correlations, say by cutting the 1D JJWs to remove the Josephson coupling on a small 
fraction of bonds, may suppress $T_c$ but is naively 
not expected to significantly impact the nematic order. 
However, an Imry-Ma argument \cite{ImryMa1975,Nie2014,Nie2017} suggests that local random fields arising from impurities
can eventually kill long-range nematic order on sufficiently long length scales. Indeed, once we step away from a regular lattice model of nematogens, but
view these as randomly located domains, such disorder is expected to be important.
Nevertheless, experiments on the (111) KTaO$_3$ 2DEG reveal a resistive anisotropy, but it is somewhat rounded compared with the sharp onset seen in our 
simulations in Fig.~\ref{fig:res}. This suggests that random field effects do not cause a complete breakdown of the nematic order on the length scale of the 
device. We suspect that homogeneous
strain fields in the device may be playing an important role in pinning the nematic order, and leading to a rounding of the nematic phase transition. 

\section{Discussion}

We have proposed a model of mesoscopic nematogens which are coupled to each other via JJWs. 
SC correlations in the JJWs have been shown to drive nematogen ordering and a spontaneous breaking of lattice rotational symmetry.
Our results explain various observations on the (111) 2DEG in KTaO$_3$, and may also be relevant to ultrathin films of Nb$_x$Bi$_2$Se$_3$ and Cu$_x$Bi$_2$Se$_3$. 
An equilibrium manifestation of nematicity in the superconductor
would be a spontaneous ellipticity in the shape of superconducting vortices, which
could be probed using a scanning superconducting quantum interference device (SQUID)~\cite{Moler1998}. Such vortices
may themselves exhibit unconventional crystal orders~\cite{Carlson2003}. 
Formulating and studying a quantum version of our model, and variants which support time-reversal breaking SC \cite{babaev1,babaev2}, are 
interesting future research directions.
Finally, we note that if
the JJWs in our model
represent dislocation lines with enhanced 1D pairing \cite{Levy2018}, and the nematogens represent the local orientation of
dislocation lines, a similar model but with randomness may be relevant for recent experiments on plastically deformed SrTiO$_3$ crystals \cite{hameed2020}.


We thank Anand Bhattacharya, Changjiang Liu, Mike Norman, Peter Littlewood, Pablo Villar Arribi, Gaurav Chaudhary, and Daniel Podolsky for fruitful discussions.
This work was supported by NSERC of Canada.
The numerical simulations were performed on the Cedar cluster, hosted by WestGrid and Compute Canada.

\bibliography{jja}

\end{document}